\documentclass[12pt]{article}

\usepackage{graphicx}
\usepackage{float}

\newtheorem{theorem}{Theorem}
\newtheorem{lemma}{Lemma}
\newtheorem{definition}{Definition}

\newtheorem{corollary}{Corollary}

\newcounter{aid}

\begin{document}

\title{Utilizing Network Structure to Bound the Convergence Rate in Markov Chain Monte Carlo Algorithms }

\author{ \vspace*{3mm} Ahmad Askarian 
\hspace*{9mm} Rupei Xu 
\hspace*{9mm} Andr\'as  Farag\'o \\
Department of Computer  Science      \\
The  University   of  Texas   at  Dallas\\
   Richardson,  Texas, U.S.A. \\
   E-mail: {\tt \{axa120431, rxx130430, farago\}@utdallas.edu} }

\date{}
\maketitle

\begin{abstract} 
\noindent
We consider the problem of estimating the measure of subsets in very large 
networks. A prime tool for this purpose is the Markov Chain Monte Carlo (MCMC)
algorithm. This algorithm, while extremely useful in many cases, still often suffers from 
the drawback of very slow convergence. We show that in a special, but important case, 
it is possible to obtain significantly better bounds on  the convergence rate. 
This special case is when the huge state space can be aggregated into a 
smaller number of clusters, in which the states behave {\em approximately}
the same way (but their behavior still may not be identical). 
A Markov chain with this structure is called {\em quasi-lumpable}. This property allows 
the {\em aggregation} of states (nodes) into clusters.
Our main contribution is a rigorously proved bound on the  rate at which the  aggregated state
distribution  approaches  its  limit  in  quasi-lumpable Markov  chains.  We  also
demonstrate numerically that in  certain cases this  can indeed lead to  a significantly
accelerated way of estimating the measure of subsets. 
The result can be a useful tool in the analysis of complex networks,
whenever they have a clustering that aggregates nodes with similar (but not necessarily identical) 
behavior.

\medskip

\noindent
%{\bf Keywords:} Markov Chain, convergence rate, lumpable
%and quasi-lumpable Markov chain.
\end{abstract}

\section{Introduction}

The Markov Chain Monte Carlo (MCMC) method is one of the most frequently used algorithms to solve hard
counting, sampling  and optimization  problems. This is relevant for  many areas,
including complex networks, physics,  communication systems, computational biology,  optimization,  data mining, big data analysis, forecast problems, prediction tasks, 
and innumerable  others. The success and  influence of the method is shown  by
the fact that it has  been  selected   as  {\em one  of   the  top  ten   of  all  algorithms
in the 20th century,} see \cite{cipra}.
The MCMC algorithm also plays an important role in large, complex networks.
In this paper, building on our earlier conference presentations \cite{farago1,farago2},  we consider the following regularly occurring
application of the MCMC method: 

Consider a very large graph $G$, with node set $S$, and let 
$A\subseteq S$ be a subset of the nodes. We would like to estimate the relative size of $A$,
that is, the goal is to obtain a good estimate of the value
\begin{equation}\label{e0}
p=\frac{|A|}{|S|}.
\end{equation}
More generally, if a random walk is considered on the graph, with stationary distribution $\pi$,
then we would  like to estimate $\pi(A)$, the stationary probability of being in $A$. In the special case when  $\pi$ is the uniform distribution, we get back the formula (\ref{e0}).

If we can take random samples from  $S$, according to the stationary distribution, then an  obvious
estimate with good properties is  the relative frequency of the  event
that the sample falls in $A$. Unfortunately, in most nontrivial  cases
of interest, this sampling  task is  not feasible.
The reason  is that  often the large  set $S$  is defined {\em implicitly}.
Examples are the  set   of all cliques in a graph,  
or the  set of all feasible solutions to an optimization  problem, and many others. No efficient general method  is
known to sample uniformly at random from such complex sets. 

An important
application in telecommunication networks is to estimate blocking probabilities, see \cite{kelly,louth}. More generally, if we have a large system, with an enormous state space, we may want to estimate that the actual state falls in a specific subset.
For example, if the state space consists of all possible load values of the network links, 
which leads to a state space of astronomical size, we may want to know what the probability is that at most $k$ links are overloaded, for some value of $k$.

At this point the MCMC does a very good service. If we define a Markov
chain in which the states are the elements of $S$ and the  transitions
are based on simple local operations, then we can very often obtain  a
Markov chain with uniform, or some other simple  stationary distribution over $S$.  Then, if
we run this chain long enough so that it gets close to the  stationary
distribution, then the state  where we stop the  chain will be a  good
approximation of a random sample over $S$, distributed according to the stationary distribution.
Then by repeating the process sufficiently many times, and by counting the relative frequency that the 
random sample falls in $A$, we can get a good estimate of the probability measure of $A$.

The key  difficulty is,  however, that  we should  run the  chain long
enough to get sufficiently close to the stationary distribution.  This
time is often referred to as {\em mixing time} \cite{sinclair}. If the
mixing time grows only polynomially with the size of the problem, e.g.
with the size of the graph, then we say that the chain is {\em rapidly
mixing}.  Unfortunately,  in  many  cases  of interest the mixing time
grows exponentially with the problem parameters, so in many  important
cases the Markov chain is mixing very slowly.

What we are interested  in is whether it  is possible to speed  up the
running time. It is clear that if we want to estimate the size of {\em
any}  possible  subset,  then  we  really  need  to  get  close to the
stationary distribution,  since only  this distribution  can guarantee
that the probability of the random state falling in the set is  really
the relative size of  the set. On the  other hand, if we  only want to
estimate the relative size  of a {\em specific} subset  $A$, then it is  enough
for us if we reach a distribution in which the measure of $A$ is close
to the stationary measure,  but this does not  have to hold for  every
other set. In other words,  if $\pi_t$ denotes the state  distribution
after $t$ steps and $\pi$ is the stationary distribution, then we want
to choose  $t$ such  that $|\pi_t(A)-\pi(A)|$  is small,  but the same
does not have to  hold for all other  sets. This makes it  possible to
reduce the required value of $t$, that is, to speed up the  algorithm.
In this paper we investigate  under what conditions it is  possible to
obtain such a speed-up. 

The main result is that the structure of the chain, that is, the 
{\em network structure}, can significantly help, if it has some special properties.
Specifically,  if the Markov chain is
close  to  a  so  called  {\em  lumpable}  chain,  then remarkable speedup is
possible. In other words, in this case we can indeed capitalize on the particular network structure
to accelerate the method.
Below we informally explain what the concept of lumpability means. 
The formal definition will follow in the next section.

%%%%%%%%%%%%%%%%%%%%%%%%%%%%%%%%%%%%%%%%%%%%%%%%%%%%%%%%%%%%%

The concept of lumpability stems from the following observation:
it is very useful if the state space can be  {\em partitioned}
such that the  states belonging to  the same partition  class ``behave
the same way," in the sense defined formally in the next section. This
is  the   concept  of  {\em  lumpability}   \cite{kemeny}.
Informally speaking, it means that  some sets of states can  be lumped
together (aggregated) and replaced by a single state, thus obtaining a Markov chain
which has  a much smaller  state space,  but its  essential behavior is the
same as the original.
In some  cases the  lumpability of  the Markov  chain can  have a very
significant effect on the efficiency of the model. A practical example
is discussed in \cite{lee,sargo}, where the authors present a fast algorithm
to compute the PageRank vector,  which is an important part  of search
engine algorithms in  the World Wide  Web. The PageRank  vector can be
interpreted as  the stationary  distribution of  a Markov  chain. This
chain has a  huge state space,  yielding excessive computation  times.
This  Markov  chain,   however,  is  lumpable.   Making  use  of   the
lumpability,  the  computation  time  can  be  reduced  to as low as
20\% of the original, according to the experiments presented in \cite{lee}.

Unfortunately, it happens relatively rarely that the Markov chain
satisfies the definition of lumpability {\em exactly}. This motivates
the concept of {\em quasi-lumpability} \cite{dayar,franc}. Informally,
a Markov chain is quasi-lumpable if its transition matrix is
obtainable by a small {\em perturbation} from a matrix that
exactly satisfies the lumpability condition (see the formal definition
in the next section).

In this  paper we  are interested  in the  following problem, which is
often encountered  in applications:  how long  do we  have to  run the
Markov chain if  we want to  get close to  the stationary distribution
within a  prescribed error?  
While the general question  is widely
discussed in  the literature  (see, e.g.,  \cite{kijima,sinclair}), we focus here 
on a less researched special case: 
how much gain can the convergence speed enjoy, if we can capitalize on the 
special structure of {\em quasi-lumpability.}

\section{Aggregation in  Markov Chains }

We assume  the reader  is familiar  with the basic concepts  of
Markov chains. We adopt the notation  that a Markov chain $\cal M$  is
given by a set  $S$ of states and  by a transition probability  matrix
$P$,  so  we  write  ${\cal  M}=(S,P)$.  This notation does not include the initial
distribution, because it is assumed arbitrary.

Let us first define the concept {\em  lumpability} of a Markov chain.  Informally,
as mentioned in  the Introduction, a  chain is lumpable  if its states
can be aggregated into larger subsets of $S$, such that the aggregated
(lumped)  chain   remains  a   Markov  chain   with  respect   to  the
set-transition probabilities (i.e, it preserves the property that  the
future depends on the past only through the present). Note that this is
generally not preserved by {\em any} partition of the state space.
Let us introduce now the formal definition.

\begin{definition} \label{deflump} {\bf (Lumpability~of~Markov chain)}  Let
${\cal M}=(S,P)$ be a Markov chain. Let ${\cal  Q}=\{A_1,\ldots,A_m\}$
be a partition  of $S$. The  chain $\cal M$  is called {\em  lumpable}
with  respect  to the partition  $\cal  Q$  if  for  any  initial  distribution  the
relationship
\begin{equation} \label{lu}
\Pr(X_t\in   A_j \, | \,
X_{t-1}\in  A_{i_1},\ldots,  X_{t-k}\in  A_{i_k})=
\Pr(X_t\in  A_j \,|\, X_{t-1}\in A_{i_1})
\end{equation}
holds for any $t, k, j, i_1, \ldots, i_k$,
whenever these conditional probabilities are defined (i.e., the
conditions occur with positive probability).
If the chain is
lumpable, then the state set of the
{\rm lumped chain} is $\cal Q$ and its state transition
probabilities are defined by $$\hat p_{ij}
=\Pr(X_t\in  A_j \,|\, X_{t-1}\in A_{i}). $$
\end{definition}

Checking whether a Markov chain is lumbable would be hard to do directly from the 
definition. That is why it is useful to have the following characterization,
which is  fundamental result on the lumpability of Markov chains, 
see \cite{kemeny}. For simple description, we use the
notation $p(x,A)$ to denote the probability that the chain moves
into a set $A\subseteq S$ in the next step, given that currently it is in the state $x\in S$.
Note that $x$ itself may or may not be in $A$.

\begin{theorem} \label{lumpthm} {\bf  (Necessary and sufficient  condition
for lumpability, see \cite{kemeny})}  A Markov  chain ${\cal  M}=(S,P)$ is  lumpable with
respect to  a partition  ${\cal Q}=\{A_1,\ldots,A_m\}$  of $S$  if and
only if for any  $i,j$ the value of  $p(x,A_j)$ is the same  for every
$x\in A_i.$  These common  values define  the transition probabilities
$\hat p(A_i,A_j)$ for the {\em lumped chain}, which is a Markov  chain
with  state  set  $\cal  Q$  and  state transition probabilities $$\hat
p(A_i,A_j)=p(x,A_j)= \Pr(X_t\in A_j  \,|\, X_{t-1}\in A_{i})  $$ where
$x$ is any state in $A_i.$ \end{theorem}

Informally, the condition means that a move from a 
set $A_i\in \cal Q$ to another set $A_j\in \cal Q$ happens with 
probability $p(x,A_j)$, no matter which $x\in A_i$ is chosen. That is, 
any $x\in A_i$ has the property that the probability of moving from this 
$x$ to the set $A_j$ in the next step is the same for every $x\in A_i$. 
The sets $A_i, A_j$ are partition classes of $\cal Q$. We also allow $i=j$,
so they may coincide.
Whenever our  Markov chain  is lumpable,  we can  reduce the number of
states by the above aggregation, and it is usually advantageous  for
faster   convergence   (a   specific   bound   will   be   proven   in
Section~\ref{main}).

It is worth noting that lumpability is a rather special property, and one has to be 
quite lucky to encounter a real-life Markov chain that actually has this 
property. Sometimes it happens (see, e.g., the example in the Introduction
about PageRank computation), but it is not very common. Therefore, let us
now relax the concept  of lumpability to broaden the family  of
the considered  Markov chains.  The extended condition, as
explained below, is called {\em quasi-lumbability.}

Informally, a  Markov chain  is called
{\em quasi-lumpable} or {\em $\epsilon$-quasi-lumpable} or simply {\em
$\epsilon$-lumpable}, if it may not  be perfectly lumpable, but it  is
``not too far"  from   that.  This   ``$\epsilon$-closeness"  is   defined  in
\cite{dayar,franc}  in  a  way  that  the  transition  matrix  can  be
decomposed  as  $P=P^-+P^\epsilon$.  Here  $P^-$  is  a  component-wise
non-negative  lower  bound  for the original transition matrix $P$,  
such  that  $P^-$  satisfies the necessary and sufficient condition of Theorem~\ref{lumpthm}. 
The other matrix, $P^\epsilon$, represents a {\em perturbation}. 
It is an arbitrary non-negative matrix in which each
entry  is  bounded  by  $\epsilon$.  This definition is not very easy to visualize, 
therefore, we use  the following simpler but equivalent definition.

\begin{definition}\label{epslump}
{\bf ($\epsilon$-lumpability)}
Let $\epsilon\geq 0.$ A Markov  chain ${\cal  M}=(S,P)$ is  called
$\epsilon$-lumpable with respect to a partition
${\cal Q}=\{A_1,\ldots,A_m\}$  of $S$  if
$$\big|p(x,A_j)-p(y,A_j)\big|\leq\epsilon $$
holds for any $x,y\in A_i$ and for any   $i,j\in\{1,\ldots,m\}$.
\end{definition}

Note  that  if  we  take  $\epsilon=0$,  then we get back the ordinary
concept of lumpability. Thus, quasi-lumpability is indeed a relaxation
of the original concept. It  can also be interpreted in  the following
way. If $\epsilon>0$, then the original definition of lumpability  may
not hold. This  means, the aggregated  process may not  remain Markov.
i.e., it does not satisfy (\ref{lu}). On the other hand, if $\epsilon$
is small, then the aggregated process will be, in a sense, ``close" to
being Markov, that is, to satisfying (\ref{lu}).
What  we  are  interested  in  is  the  convergence  analysis of quasi-lumpable Markov chains, typically for a small value of $\epsilon$. 
For the analysis we need to introduce another definition.

\begin{definition}\label{matrices}
{\bf (Lower and upper transition matrices)}
Let ${\cal  M}=(S,P)$ be a Markov chain which is
$\epsilon$-lumpable with respect to a partition
${\cal Q}=\{A_1,\ldots,A_m\}$.
The lower and upper transition matrices $L=[l_{ij}]$ and $U=[u_{ij}]$
are defined as $m\times m$ matrices with entries
$$ l_{ij} =  \min_{x\in A_i} p(x,A_j) \;\;\;\;\mbox{\rm and}\;\;\;\;
 u_{ij} =  \max_{x\in A_i} p(x,A_j), $$
respectively, for $i,j=1,\ldots,m.$
\end{definition}

Note that it always holds (component-wise) that $L\leq U$. If the
chain is lumpable, then these matrices coincide, so then $L=U=\tilde P
$, where $\tilde P$ is the transition matrix of the lumped chain. If
the chain is $\epsilon$-lumpable, then $L$ and $U$ differ at most by
$\epsilon$ in each entry.

Generally, $L$   and    $U$   are   not   necessarily    stochastic
matrices\footnote{ A vector is called stochastic if each coordinate is
non-negative and their sum is 1. A matrix is called stochastic if  each
row vector of it is stochastic.}, as  their rows may not sum up  to
1.

\section{Convergence Analysis}
\label{main}

An important concept in Markov chain convergence analysis is the {\em
ergodic coefficient}, see, e.g., \cite{kijima}. It is also called {\em coefficient of ergodicity.}

\begin{definition}\label{ergodic} {\bf (Ergodic coefficient)}
Let $P=[p_{ij}]$ be an $n\times n$  matrix. Its {\em
ergodic coefficient} is defined as
$$\rho(P)=\frac{1}{2} \max_{i,j} \sum_{k=1}^n |p_{ik}-p_{jk}|.$$
\end{definition}

The ergodic coefficient is essentially the largest $L_1$ distance that
occurs between different row vectors of the matrix $P$. That is, in a sense,
it captures how diverse are the row vectors of the matrix. The 1/2 factor is only for
normalization purposes. For stochastic matrices two important properties of the ergodic
coefficient are the following \cite{kijima}:
\begin{description}
\item[{\rm (i)}]
$\;\;\;\;0\leq \rho(P) \leq 1$
\item[{\rm (ii)}]
$\;\;\;\rho(AB) \leq \rho(A) \rho(B)$
\end{description}

The importance of the ergodic coefficient lies in its relationship  to
the convergence rate of  the Markov chain. It  is well known that  the
convergence rate is determined by the second largest eigenvalue of the
transition  matrix  (that  is,  the  eigenvalue  which has the largest
absolute  value  less  than  1),  see,  e.g., \cite{sinclair}. If this
eigenvalue  is  denoted  by  $\lambda_1$,  then  the  convergence   to
the stationary distribution happens at a rate  of $O(\lambda_1^t)$, where $t$ is  the
number  of  steps, see \cite{kijima}.  It  is  also  known \cite{kijima} that the ergodic
coefficient is always an upper bound on this eigenvalue, it  satisfies
$\lambda_1\leq  \rho(P)\leq  1$.   Therefore,  the  distance   to  the
stationary distribution is also  bounded by $O(\rho(P)^t).$ Thus,  the
smaller  is  the  ergodic  coefficient,  the faster convergence we can
expect. Of course it only provides any useful bound if $\rho(P)<1.$ If
$\rho(P)=1$ happens to be the case, then it does not directly provide a useful
bound on the convergence rate, since then $\rho(P)^t$ remains 1. In this
situation a possible way out is  considering the  $k$-step transition
matrix  $P^k$  for  some constant integer $k$.  If  $k$  is  large enough, then we can
certainly achieve $\rho(P^k)<1$, since it is known \cite{kijima}  that
$\lim_{k\rightarrow\infty} \rho(P^k) =0$.

Now we are ready to present our  main result, which is a bound on  how
fast  will  an  $\epsilon$-lumpable  Markov  chain  converge  to   its
stationary distribution on the sets that are in the partition, which is
used in defining the $\epsilon$-lumpability of the chain. We are going
to discuss the applicability of the result in the next section.

\begin{theorem}\label{converge}
Let   $\epsilon\geq   0$   and  ${\cal
M}=(S,P)$   be   an   irreducible, aperiodic Markov   chain
with stationary distribution $\pi$. Assume the chain is
$\epsilon$-lumpable    with    respect    to    a   partition   ${\cal
Q}=\{A_1,\ldots,A_m\}$ of $S$. Let $\rho$  be any upper bound
on the ergodic coefficient of the lower transition matrix $L$
{\em (Definition~\ref{matrices})}, that is, $\rho(L)\leq \rho$. Let
$\pi_0$ be any initial probability distribution on
$S$,   such   that   ${\rm   P}(X_t\in   A_i)>0$ holds  for  any  $i$,  and
$t=0,1,2,\ldots$, whenever the chain starts from $\pi_0$.
Then  for every $t\geq 1$
the     following      estimation     holds:
$$\sum_{i=1}^m\big|\pi_t(A_i)-\pi(A_i)\big|         \leq
2(\rho+\epsilon m/2)^t +
\epsilon m \frac{1-(\rho+\epsilon m/2)^t}{1-\rho-\epsilon m/2}$$
assuming $\rho+\epsilon m/2<1$.
\end{theorem}

{\em Remark:} Recall that the parameter $\epsilon$ quantifies how much the Markov chain deviates from the ideal lumpable case, see Definition 2.
In the extreme case, when $\epsilon=1$, every Markov chain satisfies the definition. This places an``upward pressure" on $\epsilon$: the larger it is, the broader is the class of Markov chains to which $\epsilon$-lumpability applies. On the other hand, a downward pressure is put on $\epsilon$ by Theorem 2:
the convergence bound is only meaningful, if $\rho+\epsilon m/2<1$ holds. This inequality can be checked for any particular $\epsilon$, 
since it is assumed that $\rho$ and $m$ are known parameters. Furthermore, the smaller is $\epsilon$, the faster is the convergence.  Therefore, the best value of $\epsilon$ is the smallest value which still satisfies Definition 2 for the considered state partition.

For  the  proof  of Theorem~\ref{converge} we  need  a  lemma  about stochastic
vectors and matrices (Lemma 3.4 in \cite{hartfiel}, see also
\cite{hartfiel2}):

\begin{lemma}\label{hartf}
{\rm (Hartfiel \cite{hartfiel,hartfiel2})}
Let $x,y$ be $n$-dimensional stochastic vectors and
$B_1,\ldots,B_k;$\,
$C_1,\ldots,C_k$ be $n\times n$ stochastic matrices. If $\rho(B_i)\leq
\rho_0$ and
$\rho(C_i)\leq \rho_0$ for all $i,\;1\leq i\leq k,$ then
$$\|xB_1\ldots B_k-yC_1\ldots C_k\| \leq
\rho_0^k\|x-y\| + \left(\sum_{j=0}^{k-1}\rho_0^j\right){\cal E}$$
%(\rho_0^{k-1}+\ldots+1){\cal E}$$
where ${\cal E}=\max_i\|B_i-C_i\|.$ The vector norm used is the
$L_1$-norm $\|x\|=\sum_{i=1}^n|x_i|$
and the matrix norm is $$\|A\|=\sup_{z\neq 0} \frac{\|zA\|}{\|z\|}=
\max_i \sum_{j=1}^n |a_{ij}|$$
for any $n\times n$ real matrix $A=[a_{ij}].$
\end{lemma}

Lemma~\ref{hartf} can be proved via induction on $k$, see
\cite{hartfiel,hartfiel2}. Now, armed with the lemma, we can prove our
theorem.

\medskip \noindent {\bf Proof of Theorem~\ref{converge}.} Let $\pi_0$
be
an  initial  state  distribution  of  the  Markov  chain $\cal M$, let
$\pi_t$  be  the  corresponding  distribution  after  $t$  steps   and
$\pi=\lim_{t\rightarrow\infty}\pi_t$ be the (unique) stationary
distribution of
$\cal  M$.  For  a  set  $A\subseteq  S$ of states the usual notations
$\pi_t(A)={\rm   P}(X_t\in   A),$\,  $\pi(A)=\lim_{t\rightarrow\infty}
\pi_t(A)$ are adopted.

Using the sets $A_1,\ldots, A_m$ of the partition $\cal Q$, let us
define the stochastic vectors
\begin{equation} \label{vector}
\tilde\pi_t=\big(\pi_t(A_1),\ldots,\pi_t(A_m)\big)
\end{equation}
for $t=0,1,2,\ldots$    and
the $m\times m$ stochastic matrices
\begin{equation} \label{matrix}
\tilde P_t(\pi_0)=
[p_t^{(\pi_0)}(i,j)]=
\big[{\rm P}(X_{t+1}\in A_j\;|\;X_t\in A_i)\big]
\end{equation}
for $t=1,2,\ldots$. Let us call them aggregated state
distribution vectors and aggregated transition matrices, respectively.
Note that although the entries in (\ref{matrix}) involve only
events of the form $\{X_t\in A_k\},$ they may also depend on the
detailed state distribution  within these sets, which is in
turn determined by the initial
distribution $\pi_0.$ In other words, if two different initial
distributions give rise to the same probabilities for the events
$\{X_t\in A_k\}$ for some $t$,  they may still result in different
conditional probabilities of the form
${\rm P}(X_{t+1}\in A_j\;|\;X_t\in A_i),$ since the chain is not
assumed lumpable in the ordinary sense.  This is why the notations
$\tilde P_t(\pi_0),\; p_t^{(\pi_0)}(i,j)$ are used.
Also note that the
conditional probabilities are well defined
for any initial distribution allowed by the
assumptions
of the lemma, since then ${\rm P}(X_t\in A_i)>0.$

For any fixed $t$ the events $\{X_t\in A_i\},\; i=1,\ldots,m,$ are
mutually exclusive with total probability 1, therefore, by the law of
total probability,
$$
{\rm P}(X_{t+1}\in A_j)=\sum_{i=1}^m
{\rm P}(X_{t+1}\in A_j\;|\;X_t\in A_i) {\rm P}(X_t\in A_i),
\;\;\;\;\;\;j=1,\ldots,m
$$
holds.
This implies
$\tilde\pi_{t+1}=\tilde\pi_t \tilde P_t(\pi_0)$, from which
\begin{equation}\label{mproduct}
\tilde\pi_t=\tilde\pi_0 \tilde P_1(\pi_0)\cdot \ldots\cdot \tilde P_t(\pi_0)
\end{equation}
follows.

We next  show that  for any
$t=1,2,\ldots$ the  matrix $\tilde  P_t(\pi_0)$ falls  between  the
lower and upper transition matrices,
i.e., $L\leq \tilde  P_t(\pi_0)\leq M$ holds.  Let us use  short
notations for certain events: for any $i=1,\ldots,m$ and for a
fixed $t\geq 1$ set $H_i=\{X_t\in A_i\},$  $H'_i=\{X_{t+1}\in A_i\},$
and   for  $x\in   S$  let   $E_x=\{X_t=x\}.$  Then   $E_x\cap
E_y=\emptyset$  holds  for  any  $x\neq  y$  and $\sum_{x\in S}E_x=1.$
Applying  the  definition  of  conditional  probability and the law of
total probability, noting that ${\rm P}(H_i)>0$ is provided by the
assumptions of the lemma, we get
\begin{eqnarray} \nonumber
p_t^{(\pi_0)}(i,j) =
{\rm P}(H'_j\:|\:H_i)
 & = & \frac{{\rm P}(H'_j\cap H_i)}{{\rm P}(H_i)}  \\ \nonumber
& = & \frac{\sum_{x\in S}{\rm P}(H'_j\cap H_i\cap E_x)}{{\rm P}(H_i)}
\\ \nonumber
& = & \frac{\sum_{x\in S}{\rm P}(H'_j\:|\: H_i\cap E_x){\rm P}(H_i\cap
E_x)}{{\rm P}(H_i)} \\ \nonumber
& = & \sum_{x\in S}{\rm P}(H'_j\:|\: H_i\cap E_x)
\frac{{\rm P}(H_i\cap E_x)}{{\rm P}(H_i)} \\ \nonumber
& = & \sum_{x\in S}{\rm P}(H'_j\:|\: H_i\cap E_x)
{\rm P}(E_x\:|\:H_i). \nonumber
\end{eqnarray}
Whenever $x\notin A_i$ we have
${\rm P}(E_x\,|\,H_i) =
{\rm P}(X_t=x\,|\,X_t\in A_i) = 0.$ Therefore, it is enough to take
the summation over $A_i$, instead of the entire $S.$ For $x\in
A_i,$ however,
$H_i\cap E_x = \{X_t\in A_i\}\cap \{X_t=x\}=\{X_t=x\}$ holds, so we
obtain
$$
p_t^{(\pi_0)}(i,j) =
\sum_{x\in A_i}{\rm P}(X_{t+1}\in A_j\:|\: X_t=x)
{\rm P}(X_t=x\:|\:X_t\in A_i).
$$
Thus,
$p_t^{(\pi_0)}(i,j)$ is a weighted average of the
${\rm P}(X_{t+1}\in A_j\,|\, X_t=x)$ probabilities. The weights
are
${\rm P}(X_t=x\:|\:X_t\in A_i),$ so they are
non-negative and sum up to 1. Further,
$$l_{ij}\leq {\rm P}(X_{t+1}\in A_j\:|\: X_t=x)\leq u_{ij}$$
must hold, since $l_{ij}, u_{ij}$ are defined as the
minimum and maximum values, respectively, of
$$p(x,A_j)=
{\rm P}(X_{t+1}\in A_j\;|\: X_t=x)$$ over $x\in A_i.$ Since the
weighted
average must fall between the minimum and the maximum, therefore, we
have
\begin{equation}\label{p}
l_{ij}\leq p_t^{(\pi_0)}(i,j) \leq u_{ij},
\end{equation}
that is,
\begin{equation}\label{LPM}
L\leq \tilde  P_t(\pi_0)\leq M
\end{equation}
for any $t\geq 1$ and for any initial distribution $\pi_0$
allowed by the conditions of the Theorem.

Let us now start the  chain from an initial distribution  $\pi_0$ that
satisfies the  conditions of  the Theorem.  We are  going to compare the
arising aggregated state distribution vectors (\ref{vector}) with  the
ones  resulting   from  starting   the  chain   from  the   stationary
distribution $\pi.$ Note  that, due to  the assumed irreducibility  of
the original chain, $\pi(x)>0$ for all  $x\in S$, so $\pi$ is also  a
possible initial  distribution  that  satisfies  the conditions
${\rm P}(X_t\in A_i)>0$.

When the  chain  is  started  from the stationary distribution $\pi$,
then, according to (\ref{mproduct}), the aggregated state distribution
vector at time $t$ is $\tilde\pi \tilde P_1(\pi)\cdot \ldots\cdot \tilde P_t(\pi)$
where $\tilde\pi$ is given as $\tilde\pi=\big(\pi(A_1),\ldots,\pi(A_m)\big).$ On the other
hand,
${\rm P}(X_t\in A_i)$ remains the same for all $t\geq 0$ if the chain
starts from the stationary distribution. Therefore, we have
\begin{equation}\label{pi}
\tilde\pi \tilde P_1(\pi)\cdot \ldots\cdot \tilde P_t(\pi) =
\tilde\pi=\big(\pi(A_1),\ldots,\pi(A_m)\big).
\end{equation}
When the chain starts from $\pi_0,$ then we obtain the aggregated
state distribution vector
\begin{equation}\label{tildepi}
\tilde\pi_t=\tilde\pi_0 \tilde P_1(\pi_0)\ldots\tilde P_t(\pi_0)
\end{equation}
after $t$ steps. Now we can apply Lemma~\ref{hartf} for the
comparison of (\ref{pi}) and (\ref{tildepi}).
The roles for the
quantities in Lemma~\ref{hartf} are assigned as $x=\tilde\pi_0,$\,
$y=\tilde\pi,$\, $k=t,$ \,$n=m,$\, and, for every
$\tau=1,\ldots,k,$\,
$B_\tau=\tilde P_\tau(\pi_0),\,$
$C_\tau=\tilde P_\tau(\pi).$
To find the value of $\rho_0$ recall that by (\ref{LPM}) we have
$L\leq \tilde  P_\tau(\pi_0)\leq M$ and
$L\leq \tilde  P_\tau(\pi)\leq M$. Since any entry of $U$
exceeds the corresponding entry of $L$ at  most by $\epsilon$,
therefore, by the definition of the ergodic coefficient,
$\rho\big(\tilde  P_\tau(\pi_0)\big)\leq\rho+\epsilon m/2$ and
$\rho\big(\tilde  P_\tau(\pi)\big)\leq\rho+\epsilon m/2$ hold, where
$\rho$ is the upper bound  on $\rho(L)$. Thus, we can take
$\rho_0=\rho+\epsilon m/2$.
With these role
assignments we obtain from Lemma~\ref{hartf}
$$
\|\tilde\pi_0 \tilde P_1(\pi_0)\ldots\tilde P_t(\pi_0) -
\tilde\pi \tilde P_1(\pi)\ldots\tilde P_t(\pi) \| \leq
(\rho+\epsilon m/2)^t\|\tilde\pi_0-\tilde\pi\| + {\cal
E}\sum_{k=0}^{t-1}(\rho+\epsilon m/2)^k
$$
where ${\cal E} = \max_\tau \|P_\tau(\pi_0)-P_\tau(\pi_0)\|$
and the norms are as in Lemma~\ref{hartf}. Taking
(\ref{pi}) and (\ref{tildepi}) into account yields
\begin{equation}\label{est}
\|\tilde\pi_t - \tilde\pi \| =
\sum_{i=1}^m\big|\pi_t(A_i)-\pi(A_i)\big| \leq
(\rho+\epsilon m/2)^t\|\tilde\pi_0-\tilde\pi\| +
{\cal E}\sum_{k=0}^{t-1}(\rho+\epsilon m/2)^k.
\end{equation}
Thus, it only remains to estimate $\|\tilde\pi_0-\tilde\pi\|$ and
$\cal E$. Given that $\tilde\pi_0, \tilde\pi$ are both stochastic
vectors, we have $\|\tilde\pi_0-\tilde\pi\| \leq
\|\tilde\pi_0\|+\|\tilde\pi\| \leq 2.$ Further,
$$
{\cal E} = \max_\tau \|P_\tau(\pi_0)-P_\tau(\pi)\| =
\max_{\tau}\, \max_i \sum_{j=1}^m
\big|p_\tau^{(\pi_0)}(i,j) - p_\tau^{(\pi)}(i,j)\big| \leq \epsilon m,
$$
since (\ref{p}) holds for any considered $\pi_0$ (including $\pi$),
and, by
the definition of $\epsilon$-lumpability, $u_{ij}-l_{ij}\leq\epsilon.$
Substituting the estimations into (\ref{est}), we obtain
\begin{eqnarray} \nonumber
\sum_{i=1}^m\big|\pi_t(A_i)-\pi(A_i)\big|    &     \leq &
2(\rho+\epsilon m/2)^t + \epsilon m\sum_{k=0}^{t-1}(\rho+\epsilon m/2)^k
\nonumber \\
& = & 2(\rho+\epsilon m/2)^t +
\epsilon m \frac{1-(\rho+\epsilon m/2)^t}{1-\rho-\epsilon m/2}
\nonumber
\end{eqnarray}
proving the Theorem.

\hfill $\spadesuit$

\medskip

If the chain happens to be exactly lumpable, then we get a ``cleaner" result. Let
$\tilde \pi_t$ be the state distribution of the lumped chain after $t$
steps and let $\tilde \pi$ be its stationary distribution. For concise
description  let  us  apply  a  frequently used distance concept among
probability  distributions.  If  $p,q$  are  two  discrete probability
distributions on  the same  set $S$,  then their  {\em total variation
distance}    $D_{TV}(p,q)$    is    defined    as    $$D_{TV}(p,q)   =
\frac{1}{2}\sum_{x\in S} |p(x)-q(x)|.$$ It  is well known that  $0\leq
D_{TV}(p,q)\leq 1$ holds for any two probability distributions. It  is
also clear from the definition  of the ergodic coefficient that  it is
the same as  the maximum total variation distance occurring between 
any two row vectors of the transition matrix.
Note   that    exact   lumpability    is   the    special   case    of
$\epsilon$-lumpability  with  $\epsilon=0$.  Therefore, we immediately
obtain the following corollary.

\begin{corollary} If the Markov chain in Theorem~\ref{converge} is  exactly
lumpable,  then  in  the  lumped  chain  for  any $t=0,1,2,\ldots$ the
following holds: $$D_{TV}(\tilde\pi_t,\tilde\pi)\leq \rho^t$$
where $\rho=\rho(\tilde P)$ is the ergodic coefficient of the
transition matrix $\tilde P$ of the lumped chain.
\end{corollary}

\noindent
{\bf Proof.} Take the special case $\epsilon=0$ in Theorem~\ref{converge}.

\hfill $\spadesuit$

\section{Numerical Demonstration}

Let us  consider the  following situation.  Let $\cal  M$ be  a Markov
chain with  state space  $S$. Assume  we want  to estimate the
stationary measure $\pi(A)$  of a subset  $A\subseteq S$. A  practical
example of such a situation is to estimate the probability that  there
is at most $k$  blocked links, for some constant $k$,  in a large communication 
network. Here the state space  is the set $S$ of all possible states of all the links. The state of a link is the current traffic load of the link, and it is blocked if the 
load is equal to the link capacity, so it cannot accept more traffic. Within this state space the considered subset $A$ is the subset of states in which among all links at most $k$ are blocked. Therefore, the relevant partition of $S$ is $\{A, S-A\}$. This is motivated by real-world application, since the number of blocked links critically affects network performance.
When considering the loss of traffic due to blocking, the models
of these networks are often called  {\em loss  networks}. For detailed 
background information  on
loss  networks,  see  \cite{kelly,mazumdar,ross}.  Of  course,  we  can also
consider other events in the network. For example, at most a given percentage of traffic  is
blocked, without specifying how many links are involved in the blocking.

In many cases  we are unable  to directly compute  $\pi(A)$. This task
frequently has enormous complexity, for the theoretical background see
\cite{louth}. Then a natural way  to obtain an estimation of  $\pi(A)$
is simulation. That is, we run the chain from some initial state, stop
it after $t$ steps and check out whether the stopping state is in  $A$
or not. Repeating this experiment a large enough number of times,  the
relative frequency of ending up in $A$ will give a good estimation  of
the measure of $\pi_t(A)$. If  $t$
is  chosen  such  that  $\pi_t$  is  close  enough  to  the stationary
distribution $\pi$ for any initial  state, then we also obtain  a good
estimation for $\pi(A)$. This is the core idea of the Markov Chain
Monte Carlo approach.

Unfortunately, Markov chains with large state space often converge
extremely  slowly. Therefore, we may not get close enough to $\pi$
after a reasonable number of steps. In such a case our result can do a
good service, at least when the chain satisfies some special
requirements. As an example, let us consider the following case. 
First we examine it using our bounds, then we also study it through numerical
experiments.

Assume the  set $A\subseteq  S$ has the property that its elements behave similarly 
in  the following sense: for
any state $x\in A$ the probability to move out of $A$ in the next step,
given that the current state is $x$, 
is {\em approximately} the same. Similarly, if $x\notin A$, then
moving into $A$ in the next step from the given $x$ has approximately the same
probability for any 
$x\notin A$. To make this assumption formal, assume there are values
$p_0,q_0, \epsilon$, such that the following conditions hold:
\begin{description}
\item[\rm (A)]
If $x\in A$ then \,\, $p_0\leq p(x,\bar A)\leq p_0+\epsilon$\, where
$\bar A =S-A$. This means, the smallest probability of moving out of $A$ from any state in $x\in A$ is at least $p_0$, and the largest such probability is at most  $p_0+\epsilon$.

\item[\rm (B)]
If $x\in \bar A$ then \,\, $q_0\leq p(x, A)\leq q_0+\epsilon$.
Similarly to the first  case, this means that the smallest probability of moving into $A$ from any state in $x\notin A$ is at least $q_0$, and the largest such probability is at most  $q_0+\epsilon$.
(We choose $\epsilon$ such that it can serve for this purpose in both directions.)

\item[\rm (C)]
To avoid degenerated cases, we require that the numbers $p_0,q_0, \epsilon$ satisify
$ p_0+\epsilon<1$,\,\,\,
$q_0+\epsilon<1$ and $0<p_0+q_0<1.$ 
The other state transition probabilities (within $A$ and $\overline A$) can be completely arbitrary, assuming, of course,  
that at any state the outgoing probabilities must sum up to 1. 

\end{description}

Let us now apply our main result, Theorem~\ref{converge}, for this situation. The parameters
will be as follows: $m$, the number of sets in the partition, is 2, since the partition is 
$(A,\overline A)$. The matrices $L, U$ become the following:
$$
L= \left[
    \begin{array}{cc}
      1-p_0-\epsilon  &  p_0 \\
      q_0             &  1-q_0-\epsilon
    \end{array} \right]
\;\;\;
\;\;\;
\;\;\;
U= \left[
    \begin{array}{cc}
      1-p_0  &  p_0+\epsilon \\
      q_0+\epsilon   &  1-q_0
    \end{array} \right].
$$
Furthermore, we can take $\rho=1-p_0-q_0-\epsilon$ as an upper bound on the ergodic coefficient of $L$.
 Then we obtain from Theorem~\ref{converge}, expressing the estimation in terms of the
total variation distance:
 $$D_{TV}(\tilde\pi_t,\tilde\pi)\leq
(1-p_0-q_0)^t +
\epsilon  \frac{1-(1-p_0-q_0)^t}{p_0+q_0} $$
where the distributions $\tilde\pi_t,\tilde\pi$ are over the sets of
the partition ($A,\bar A$), not on the original state space. Note that
in our case we actually have
 $D_{TV}(\tilde\pi_t,\tilde\pi)= |\pi_t(A)-\pi(A)|$, due to the fact
that  $|\pi_t(A)-\pi(A)|=|\pi_t(\bar A)-\pi(\bar A)|$.
Therefore, we obtain the estimation directly for the set $A$:
\begin{equation}\label{bound1}
|\pi_t(A)-\pi(A)| \leq
(1-p_0-q_0)^t +
\epsilon  \frac{1-(1-p_0-q_0)^t}{p_0+q_0}. 
\end{equation}
 If $p_0+q_0$ is
not extremely small, then the term
$(1-p_0-q_0)^t$
will quickly vanish, as it approaches 0 at an exponential rate. 
Therefore, after a reasonably small number of steps, we reach a
distribution $\pi_t$ from {\em any} initial state, such that approximately
the following bound is satisfied:
\begin{equation}\label{bound2}
|\pi_t(A)-\pi(A)| \leq
 \frac{\epsilon}{p_0+q_0}. 
\end{equation}

%\noindent 
It is quite interesting to note that neither the precise estimation (\ref{bound1}), nor its 
approximate version (\ref{bound2}) depend on the size of the state space.

Now we demonstrate via numerical results that the obtained bounds indeed hold. Moreover, they are achievable 
after a small number of Markov chain steps, that is, with fast convergence. 
We simulated the example with the following parameters: $n=100$ states, $p_0=q_0=0.25$, and $\epsilon =0.1$.
The set $A$ was a randomly chosen subset of 50 states. The transition probabilities were also chosen randomly, with the 
restriction that together with the other parameters they had to satisfy conditions (A), (B), (C).

Figure 1 shows the relative frequency of visiting $A$, as function of the number of Markov chain steps. 
It is well detectable  that the chain converges quite slowly. Even after many iterations the deviation from the stationary 
probability $\pi(A)$ does not visibly tend to 0. On the other hand, it indeed stays within our error bound:
$$
|\pi_t(A)-\pi(A)| \leq
 \frac{\epsilon}{p_0+q_0}= \frac{0.1}{0.25+0.25} = 2\cdot 0.1,
$$
as promised. Having observed this, it is natural to ask, how soon can we reach this region, that is, how many steps are needed
to satisfy the bound? This is shown in Figure 2. We can see that after only 10 iterations, the error bound is already satisfied. Note that
this is very fast convergence, since the number of steps to get within the bound was as little as 10\% of the number of states.

\medskip\medskip

\begin{figure}[H]
\begin{center}
  \includegraphics[width=12cm]{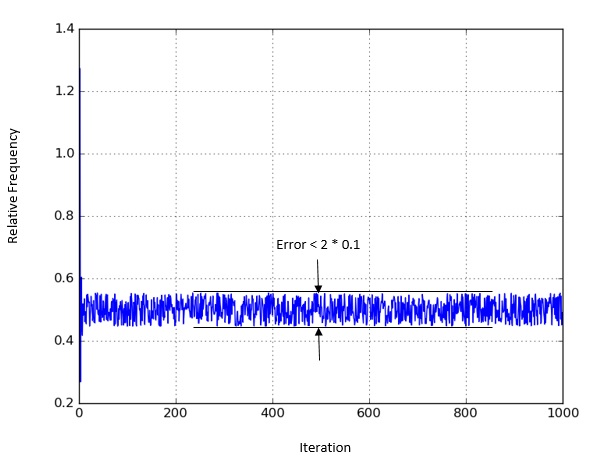}
  \caption{Deviation from the stationary measure for many iterations.}
  \label{fig:fig1}
\end{center}
\end{figure}

\begin{figure}[H]
\begin{center}
  \includegraphics[width=12cm]{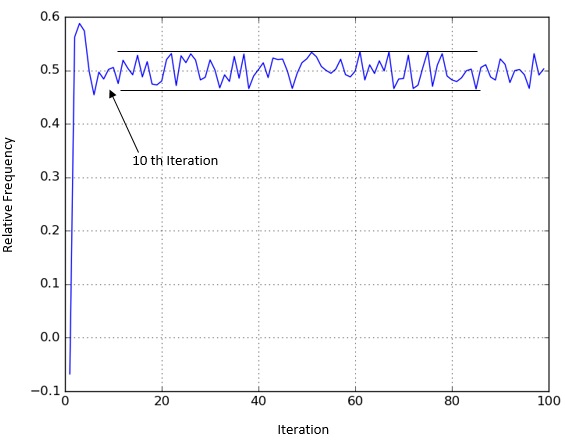}
  \caption{Very fast convergence to satisfy the error bound.}
  \label{fig:fig2}
\end{center}
\end{figure}

\section{Conclusion}

We have analyzed the convergence rate of quasi-lumpable Markov Chains.
This represents the case when the large state space can be aggregated into a 
smaller number of clusters, in which the states behave {\em approximately}
the same way.
Our main contribution is a  bound on the  rate at which the  aggregated state
distribution  approaches  its  limit  in  such  chains.  We  have also
demonstrated that in  certain cases this  can lead to  a significantly
accelerated convergence to an approximate estimation of  the measure of subsets. 
The result  can serve as a useful tool in the analysis of complex networks,
when they have a clustering that approximately satisfies the conditions lumpability.

\end{document}